\documentclass[11pt,letterpaper]{article}
\usepackage[margin=1in]{geometry}
\usepackage[T1]{fontenc}
\usepackage{lmodern,microtype,amsmath,amssymb,amsthm,booktabs,array,graphicx}
\DeclareMathSizes{10}{10}{7}{7}
\usepackage[table]{xcolor}
\definecolor{MidnightBlue}{RGB}{25,55,85}
\usepackage[colorlinks=true,linkcolor=MidnightBlue,citecolor=MidnightBlue,urlcolor=MidnightBlue]{hyperref}
\hypersetup{pdftitle={Curvature Blindness: Material Inference in Stars and Black Hole Candidates},pdfauthor={Jacob M. Krebs}}
\usepackage{fancyhdr}
\newtheorem{proposition}{Criterion}
\theoremstyle{definition}
\newtheorem{definition}{Definition}
\newtheorem{distinction}{Distinction}
\newtheorem*{evolutionremark}{Remark (shared evolution)}
\newcommand{\tot}{\mathrm{total}}
\title{\textbf{Curvature Blindness:}\\[5pt]\large Material Inference in Stars and Black Hole Candidates}
\author{Jacob M. Krebs\\\small Arkansas Center for Space and Planetary Sciences\\\small University of Arkansas}
\date{}
\begin{document}
\maketitle
\begin{abstract}
Gravitational observations can leave material properties undetermined in two different ways: distinct states can have the same total source, and distinct spacetime histories can have the same retained observations. We organize these inverse problems in general relativity through a property-level identifiability criterion, with applications to stars and black hole candidates. Two ideal massless gases illustrate equal-source ambiguity: their temperatures differ by a factor of $(8/29)^{1/4}$ and their entropies differ while their stress--energy tensors agree. A separate causal-fluid example starts with identical sources and geometric data but different constitutive responses; the two models' curvature histories separate at linear order in proper time and their scale factors at cubic order. For black hole candidates, the relevant distinction is between stationary exterior parameters, perturbing fluxes and constituent history. The framework provides a test for whether retained data identify a material property, separating instantaneous source equality, history-wide source equality and observational projection.
\end{abstract}

\section{From a gravitational description to a physical object}

Gravitational and material descriptions answer different questions about a compact object. Observations constrain its gravitational field and response; a material model specifies composition, stresses, reactions and transport. We distinguish what observations establish (the measured geometric and response properties) from the material properties compatible with them. We call the underdetermination of material properties by specified geometric data \emph{curvature blindness}. Its scope depends on the physical models and observations being compared; response measurements can supply information absent from an instantaneous source.

Here, a \emph{black hole candidate} is an observed compact object whose measured properties are consistent with a black hole description. This observational designation does not by itself specify a material interior. When discussing uniqueness or the stationary first law, \emph{black hole} instead denotes the spacetime satisfying the hypotheses of those results. By \emph{constituents}, we mean particles or material components, such as electrons, nucleons, quarks and radiation. Their composition, state and interactions belong to a material model. The application to black hole candidates is central: exterior agreement constrains a proposed object, while its material realization requires a complete source and coupled evolution. For mathematical black holes, the corresponding inverse question concerns constituent histories compatible with the retained exterior data, as developed in Sec.~\ref{sec:nohair}.

In stars, electron degeneracy, neutron statistics and nuclear interactions enter through the equation of state~\cite{chandra,ov,prakash}; temperature, opacity and reaction rates determine transport and evolution. The gravitational equations receive the total stress--energy tensor. Reconstructing its material origin is an inverse problem.

Persistence and material support also ask different questions. For a conventional black hole model, persistence concerns the stability of its exterior spacetime, not a hydrostatic interior. A proposed material object instead requires a structure and stability calculation. Comparisons of such candidates already use response, absorption and emission as well as background geometry~\cite{cardoso}. Table~\ref{tab:questions} summarizes the shared questions and their different observational access.

\begin{table}[ht]
\centering\small
\rowcolors{2}{black!5}{white}
\begin{tabular}{>{\centering\arraybackslash}m{0.21\linewidth}>{\centering\arraybackslash}m{0.35\linewidth}>{\centering\arraybackslash}m{0.35\linewidth}}
\toprule
Physical question & Stars, white dwarfs and neutron stars & Black hole candidates\\
\midrule
Gravitational field & Constrained by orbital and other available probes & Constrained by exterior and response probes\\[3pt]
Persistence & Material structure and stability & Exterior stability for a black hole model; material stability for an alternative\\[3pt]
Transport and emission & Determined by the state, interactions and boundary conditions & Require the corresponding material and radiative account\\[3pt]
Formation and evolution & Depend on composition, initial data and environment & Require a history, not only final exterior parameters\\[3pt]
Model discrimination & Uses spectra, radii, oscillations and other responses & Uses accessible multipoles, tides, absorption and emitted signals\\
\bottomrule
\end{tabular}
\caption{Physical questions shared across compact-object studies, with different model conditions and observational access. Material support and exterior persistence are distinct tests.}
\label{tab:questions}
\end{table}

The organizing question is whether all physical models compatible with specified observations agree on the material property being inferred. Equal-source ambiguity and the loss of information when a full history is reduced to exterior observations are distinct ways that this test can fail. Sections~3--4 illustrate the first mechanism and its response dependence; Sections~5--6 apply the same inference test to exterior parameters and fluxes.

The source, evolution and observation maps of Sec.~4 distinguish instantaneous source equality, history-wide source equality and equality of retained observations. Applied to black hole candidates, they separate inference of a material origin from the forward construction of an object: admissible material states and their interactions must supply the complete source, response and coupled evolution. The examples serve as worked controls for these distinctions.

Observational underdetermination in relativity has an established literature. Malament distinguishes several strengths of observational indistinguishability, each extendable to comparisons that also preserve matter fields; Manchak studies limitations on determining global spacetime structure from observable pasts~\cite{malament,manchak}. Those causal-access questions differ from both the decomposition of a locally specified total source and the restriction to selected observational summaries used here. Specifying the observation map keeps these questions separate.

\section{Connecting the two descriptions}

The material state supplies the total stress--energy tensor in the Einstein equation:
\begin{equation}
\boxed{\begin{gathered}
G_{ab}=\frac{8\pi G}{c^4}T^{\tot}_{ab},\\[3pt]
T^{\tot}_{ab}=T^{\mathrm{matter}}_{ab}+T^{\mathrm{radiation}}_{ab}
+T^{\mathrm{fields}}_{ab}+T^{\mathrm{interactions}}_{ab}+\cdots .
\end{gathered}}
\label{eq:einstein}
\end{equation}
Unless a component is explicitly indicated, $T_{ab}$ below denotes this total source. The partition on the right is schematic: radiation is itself a field, and interaction energy must be counted once within a consistent description. For example, a fluid coupled to electromagnetism can be organized into material and electromagnetic stresses, with their exchange fixed by the coupled equations. The first law expresses energy conservation in thermodynamic form, relating changes in material energy to heat, work and changes in constituent numbers:
\begin{equation}
\boxed{dE=T\,dS-P\,dV+\sum_i\mu_i\,dN_i+\Omega_{\rm mat}\,dJ+\Phi_{\rm mat}\,dQ+\cdots .}
\label{eq:firstlaw}
\end{equation}
Equation~\eqref{eq:firstlaw} is schematic: rotation and charge variations enter when independent variables of the material system, with conjugates $\Omega_{\rm mat}$ and $\Phi_{\rm mat}$. The material model supplies constitutive relations for stresses, currents and transport from material variables; together with reaction laws, these close the evolution equations. We use signature $(-,+,+,+)$ and, after the dimensional Einstein equation, set $G=c=\hbar=k_B=1$ unless stated otherwise. The label MODEL marks assumption-dependent worked controls.

The forward problem evolves specified matter with gravitational initial and boundary data. The inverse problem starts with observations and asks which coupled physical models fit them.

The connection can be written explicitly for a homogeneous material element in local equilibrium, with no independent rotation or gauge-charge variation. Let $V$ be its proper rest-frame volume, $s=S/V$ its entropy density and $n_i=N_i/V$ its independent constituent densities. Its rest-frame material energy, including interactions, is $E=V\rho(s,n_i)$. Differentiating gives
\begin{equation}
dE=\rho_{,s}\,dS+\sum_i\rho_{,n_i}\,dN_i
+\left(\rho-s\rho_{,s}-\sum_i n_i\rho_{,n_i}\right)dV.
\label{eq:energybridge}
\end{equation}
Comparing with the local first law identifies
\begin{equation}
T=\rho_{,s},\qquad \mu_i=\rho_{,n_i},\qquad
P=sT+\sum_i n_i\mu_i-\rho
=-\left(\frac{\partial E}{\partial V}\right)_{S,N_i}.
\end{equation}
For an isotropic perfect fluid these quantities supply the full source~\cite{andersson}:
\begin{equation}
T_{ab}=(\rho+P)u_a u_b+P g_{ab},\qquad u_a u^a=-1.
\end{equation}
Thus the material energy and its response to changes of state determine the energy density and stresses entering $T_{ab}$. Interaction contributions enter through $\rho$ and its derivatives. Anisotropic stress and transport require their corresponding material variables and constitutive relations. This local extensive construction uses a material element with negligible boundary contributions; gravitational initial and boundary data belong to the coupled spacetime problem.

Temperature, chemical potentials and pressure arise together from the material energy function, subject to reaction, conservation and equilibrium constraints. A proposed decomposition of the source must satisfy those relations.

Kunz uses total-source ambiguity to formulate dark-component degeneracy~\cite{kunz}. The ideal-gas example below applies this inference issue to temperature, entropy and species content.

Statistics affects the occupations, energy density and pressure through the material theory; the Einstein equation receives their total tensor.

Given a full metric and the specified field equations, Eq.~\eqref{eq:einstein} fixes the total tensor, leaving the question of which material descriptions realize it.

The inverse stellar structure problem supplies a positive example: within a specified stellar class, mass--radius information can recover a barotropic EOS~\cite{lindblom92,lindblom}. Alford et al. exhibit hybrid and nucleonic stars with very similar mass--radius relations~\cite{alford}, while relativistic multi-fluid theory retains separate currents and interactions~\cite{andersson}. Identifiability therefore depends on which property is asked for: recovering an EOS and recovering its microscopic composition are different tasks.

Borsboom and Visser develop equilibrium thermodynamic representations, equations of state and response functions for a declared black hole cavity system~\cite{borsboom}. Constituent reconstruction asks a different question: which material states are compatible with geometric data? Such an equilibrium representation does not by itself specify the constituent description used to test that compatibility.

\section{Same source, different physical states}

Different constituent populations can share a total stress--energy tensor even when their temperatures and entropies differ. This is an instance of the thermodynamic inverse problem studied by Coll, Ferrando and S\'aez~\cite{coll}: determine which thermodynamic fluids can realize a given perfect-fluid source history. Their radiation-fluid family already permits different temperature and entropy normalizations at fixed source. The following degree-counting example makes that ambiguity explicit for specified ideal species.

MODEL: An elementary construction makes the inverse question explicit. Consider ideal, massless gases with isotropic thermal distributions at zero chemical potential. We specify a common temperature for the components of each mixture; preparing and equilibrating that state is a separate material problem. In units $\hbar=c=k_B=1$,
\begin{equation}
\rho=\frac{\pi^2}{30}g_*T^4,\qquad P=\frac{\rho}{3},\qquad
T_{ab}=(\rho+P)u_a u_b+P\,g_{ab},
\label{eq:gas}
\end{equation}
with signature $(-,+,+,+)$ and $u_a u^a=-1$. Here $g_*$ includes the fermionic statistical factor $7/8$.

For an exact model comparison, let gas A have two massless bosonic degrees of freedom, $g_{*,A}=2$, and let gas B add six massless fermionic degrees of freedom, $g_{*,B}=2+(7/8)6=29/4$. At common four-velocity, choose
\begin{equation}
T_B=\left(\frac{8}{29}\right)^{1/4}T_A.
\label{eq:temperature}
\end{equation}
Then $\rho_B=\rho_A$, $P_B=P_A$, and the entire tensor in Eq.~\eqref{eq:gas} is identical. The temperatures, particle populations and entropies differ. In particular,
\begin{equation}
\frac{T_B}{T_A}=\left(\frac{8}{29}\right)^{1/4}\simeq0.724725,\qquad
\frac{s_B}{s_A}=\left(\frac{29}{8}\right)^{1/4}\simeq1.379835,
\label{eq:ratios}
\end{equation}
where $s=4\rho/(3T)$.

The degree counts correspond to photons and three neutrino/antineutrino flavors in an idealized ultrarelativistic description. For physical neutrinos, neglecting mass requires $k_BT\gg m_\nu c^2$ for each populated species, including at $T_B$. The common thermal distribution and zero chemical potentials are stipulated here. They do not specify a physical equilibration mechanism or a complete thermal plasma: other species participating in equilibration must be included in a physical application. Finite masses, interactions and nonthermal distributions likewise change the source calculation.

This is an exact background degeneracy within the specified ideal, massless description. Both descriptions also have the same barotropic derivative $dP/d\rho=1/3$. If their perturbations are closed solely as this perfect fluid, they share the corresponding gravitational--fluid perturbation problem with matched data and boundary conditions. Their responses may become distinguishable through differences in transport regime between species, energy-dependent transport or differing anisotropic stresses. If all components instead undergo the same collisionless massless transport with matched total energy-weighted angular distributions and boundary data, free streaming alone need not distinguish them. This differs from a mixture in which photons remain fluid-like while neutrinos free-stream.

For a static spatial example, impose the same boundary traction on both finite volumes. Any confining apparatus contributes to the total source and must have matching stress--energy and boundary data in both constructions. The hydrostatic interiors then coincide, as does any common compatible completion. Appendix~\ref{app:control} gives the boundary assumptions and numerical checks.

Within a material class admitting both degree counts, temperature and species content are therefore not identifiable from the total tensor. Additional material measurements can distinguish the two states. Equality of their sources throughout an evolving region is conditional on their shared perfect-fluid closure, matched full data and boundary conditions; the equality at the specified initial state alone does not impose those conditions.

\section{What the observations determine}

The same test can be used for stellar data or black hole exterior data: specify the physical class, identify all states compatible with the observations, and ask whether those states agree on the property of interest.

\begin{definition}[Physical models and observational map]\label{def:model}
The admissible class $\mathcal A$ consists of physical models $\lambda$ satisfying specified conditions. Each model includes its material variables and closure, constraint-satisfying gravitational initial data and boundary conditions, and the history and observation epoch needed for the question. For observation space $\mathcal O$, the coupled forward problem defines
\begin{equation}
F:\mathcal A\longrightarrow\mathcal O,\qquad
\lambda\longmapsto\hbox{specified observations}.
\end{equation}
For data $o$ in the image of $F$, the exact admissible inverse is
\begin{equation}
\mathcal A_o=\{\lambda\in\mathcal A:F(\lambda)=o\}.
\label{eq:fiber}
\end{equation}
This set contains all allowed models matching the observations. The property $q:\mathcal A\to\mathcal Q$ assigns a value to each model, such as temperature, composition or an initial inventory.
\end{definition}

The observations may specify the complete metric modulo gauge or a smaller set of measurements. These choices define different inverse problems. Lensing and orbits, for example, probe metric information even in vacuum regions where $G_{ab}=0$. The global identifiability question is whether all models with the same data agree on $q$.

Criterion~\ref{crit:identifiability} is the identifiability of a structural characteristic in the sense of Koopmans and Reiers\o l~\cite{koopmans}, stated for a deterministic forward map. Related parameter-identifiability formulations appear in Refs.~\cite{bellman,rothenberg}.

\begin{proposition}[Property identifiability]\label{crit:identifiability}
A property $q:\mathcal A\to\mathcal Q$ is identifiable from the declared observations if and only if it is constant on every nonempty inverse set $\mathcal A_o$. Equivalently, there exists a function $\widetilde q:F(\mathcal A)\to\mathcal Q$ that reads the property from the observations, such that $q=\widetilde q\circ F$.
\end{proposition}
Indeed, a function of the observations is constant on every inverse set. Conversely, if $q$ is constant there, assigning that common value to each observation defines $\widetilde q$.

At a particular model $\lambda_0$, the question is whether $q$ is constant on $F^{-1}(F(\lambda_0))$, rather than on every inverse set~\cite{koopmans}. This is the form used when comparing histories compatible with one specified final exterior.

For observations with a stated error model, replace exact equality by the corresponding compatibility condition. One convenient form is
\begin{equation}
\mathcal A_{o,\epsilon}=\{\lambda\in\mathcal A:
d(F(\lambda),o)\leq\epsilon\},
\end{equation}
where $d$ and $\epsilon$ are specified by the measurement model. The inferred range is $q(\mathcal A_{o,\epsilon})$. Additional material evidence restricts $\mathcal A$ and can shrink this range. A probability distribution over the compatible states requires a declared statistical measure in addition to the compatibility set.

\begin{distinction}[Equal-source and projection degeneracy]\label{dist:degeneracy}
\emph{Instantaneous equal-source degeneracy} occurs when distinct material states give the same $T_{ab}$ on an initial slice. \emph{History-wide equal-source degeneracy} requires the tensors to remain equal throughout a common evolving spacetime region. The latter depends on the material equations and full data as well as the initial source. \emph{Projection degeneracy} occurs when distinct full histories give the same retained observations. It depends on which observations are kept.
\end{distinction}

\begin{figure}[!ht]
\centering
\includegraphics[width=\linewidth]{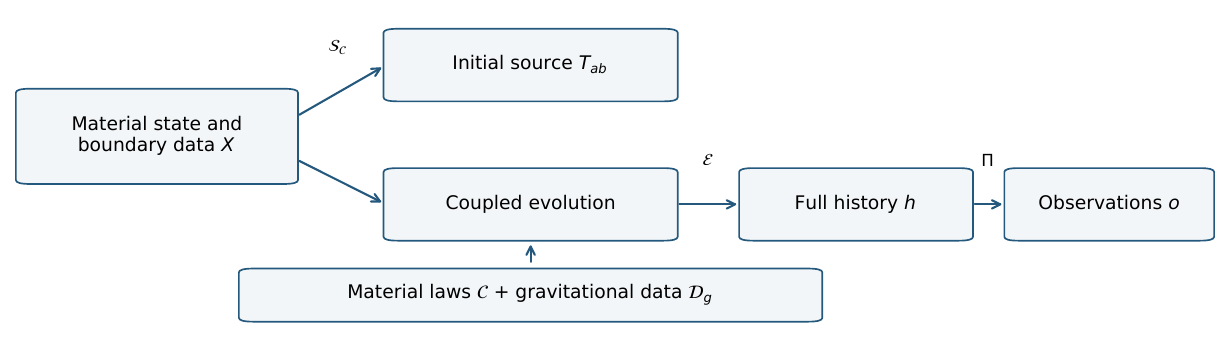}
\caption{The material state enters both source construction, evaluated on the given initial metric, and coupled evolution. Equal initial sources do not remove the need for the full material state, material laws, and gravitational data in the evolution problem.}
\label{fig:stages}
\end{figure}

These stages can be separated explicitly. Let $X$ specify the material initial and boundary data, $\mathcal C$ the constitutive, reaction and transport laws, and $\mathcal D_g$ the gravitational initial and boundary data. Then
\begin{equation}
T_{ab}(t_0)=\mathcal S_{\mathcal C}(X;g_0),\qquad
h=\mathcal E(X,\mathcal C,\mathcal D_g),\qquad o=\Pi(h),
\label{eq:stages}
\end{equation}
where $g_0$ is the initial metric included in $\mathcal D_g$, and $h$ is the full coupled history on a specified domain. The material state feeds both the source calculation and the evolution (Fig.~\ref{fig:stages}). Equal-source degeneracy concerns non-injectivity of the source map over the declared class at the given metric. Its persistence is decided by $\mathcal E$, while $\Pi$ determines which differences in histories are observable. For $\lambda=(X,\mathcal C,\mathcal D_g)$, the observational map is $F(\lambda)=\Pi[\mathcal E(X,\mathcal C,\mathcal D_g)]$.

\begin{evolutionremark}
Models with the same constraint-satisfying initial data, boundary conditions and closed gravitational--material evolution equations share that evolution, modulo gauge, within the domain where the system has a unique solution. All material variables required by the constitutive and reaction laws must match; equality of the initial $T_{ab}$ alone is insufficient. The Cauchy-development framework of Choquet-Bruhat and Geroch~\cite{cbg} provides the gravitational setting; the chosen matter equations must supply their own well-posed evolution.
\end{evolutionremark}

\subsection{Where a material assumption changes the evolution}\label{sec:evolution}

Equal sources at one epoch need not remain equal. Material responses determine how the source changes under compression, reactions or transport. A wrong constituent assumption can therefore be invisible in the initial geometry and become testable when it predicts the wrong response. Coll and Ferrando's hydrodynamic characterization of local thermal equilibrium underlies the thermodynamic inverse problem developed by Coll, Ferrando and S\'aez~\cite{coll1989,coll}. Their reconstruction distinguishes barotropic and non-barotropic evolutions. The barotropic example below instead specifies its response directly at fixed entropy per particle, as realized in Appendix~\ref{app:evolution}. Visser relates cosmological equation-of-state derivatives to higher time derivatives of the scale factor~\cite{visserjerk}; the example displays the corresponding separation in curvature and scale factor. The ideal radiation gases of Sec.~3 retain identical perfect-fluid evolution under their common closure; differences of temperature and entropy alone do not supply a difference of evolution.

MODEL: To isolate the response dependence, consider two barotropes through the same state $(\rho_0,P_0)$:
\begin{equation}
P_i(\rho)=P_0+\chi_i(\rho-\rho_0),\qquad
0<\chi_i<1,\qquad i=A,B.
\label{eq:responsecontrol}
\end{equation}
Here $\chi_i=dP_i/d\rho$ is the squared sound speed along the specified fixed-entropy-per-particle response, in the stated units. This is a local fluid-model control, restricted to a neighborhood where the energy density and enthalpy are positive and $|P_i|\leq\rho$. It supplies different causal responses without assigning them to particular microscopic species.

Write both metrics in the same cosmic proper-time coordinates:
\begin{equation}
\begin{aligned}
ds_A^2&=-dt^2+a_A(t)^2(dx^2+dy^2+dz^2),\\
ds_B^2&=-dt^2+a_B(t)^2(dx^2+dy^2+dz^2),\\
a_A(t_0)&=a_B(t_0)=1,\qquad
\dot a_A(t_0)=\dot a_B(t_0)=H_0,\qquad
H_0^2=\frac{8\pi}{3}\rho_0 .
\end{aligned}
\label{eq:initialmetric}
\end{equation}
Choose the same nonzero sign of $H_0$ for expansion or contraction. Both models have $P_i(t_0)=P_0$, so their initial spatial metric, expansion and source agree. Equality of metric values on one slice would not by itself establish equal curvature. Here the common pressure also fixes
\begin{equation}
\ddot a_i(t_0)=-\frac{4\pi}{3}(\rho_0+3P_0).
\label{eq:initialacceleration}
\end{equation}
In a common comoving orthonormal frame, the independent initial Riemann components are given below, with components containing exactly one time index vanishing:
\begin{equation}
\begin{aligned}
R_{\hat 0\hat j\hat 0\hat k}(t_0)
&=\frac{4\pi}{3}(\rho_0+3P_0)\,\delta_{jk},\\
R_{\hat j\hat k\hat l\hat m}(t_0)
&=\frac{8\pi}{3}\rho_0
 (\delta_{jl}\delta_{km}-\delta_{jm}\delta_{kl}).
\end{aligned}
\label{eq:initialcurvature}
\end{equation}
In FLRW the Weyl tensor vanishes, so the Riemann tensor is fixed by the Ricci tensor and hence by $\rho$ and $P$ through the Einstein equation. Its initial equality here follows from the symmetry and matched data; it is not an independent restriction on the two models. In a general spacetime, equal $T_{ab}$ fixes the Ricci part but leaves Weyl curvature to the gravitational data. These expressions use the Riemann convention for which $\mathcal R=6(\ddot a/a+H^2)$. The subsequent coupled equations are
\begin{equation}
H_i^2=\frac{8\pi}{3}\rho_i,\qquad
\dot\rho_i=-3H_i(\rho_i+P_i),\qquad
\mathcal R_i=8\pi(\rho_i-3P_i).
\label{eq:evolutioncontrol}
\end{equation}
Here $H_i=\dot a_i/a_i$, $\mathcal R_i$ is the Ricci scalar, and the dot is cosmic proper-time differentiation. The initial density changes agree, but the initial pressure changes depend on $\chi_i$. Consequently, with $\Delta=B-A$,
\begin{equation}
\left.\Delta\dot{\mathcal R}\right|_{t_0}
=72\pi H_0(\rho_0+P_0)(\chi_B-\chi_A).
\label{eq:curvaturedeparture}
\end{equation}

\begin{figure}[!ht]
\centering
\includegraphics[width=\linewidth]{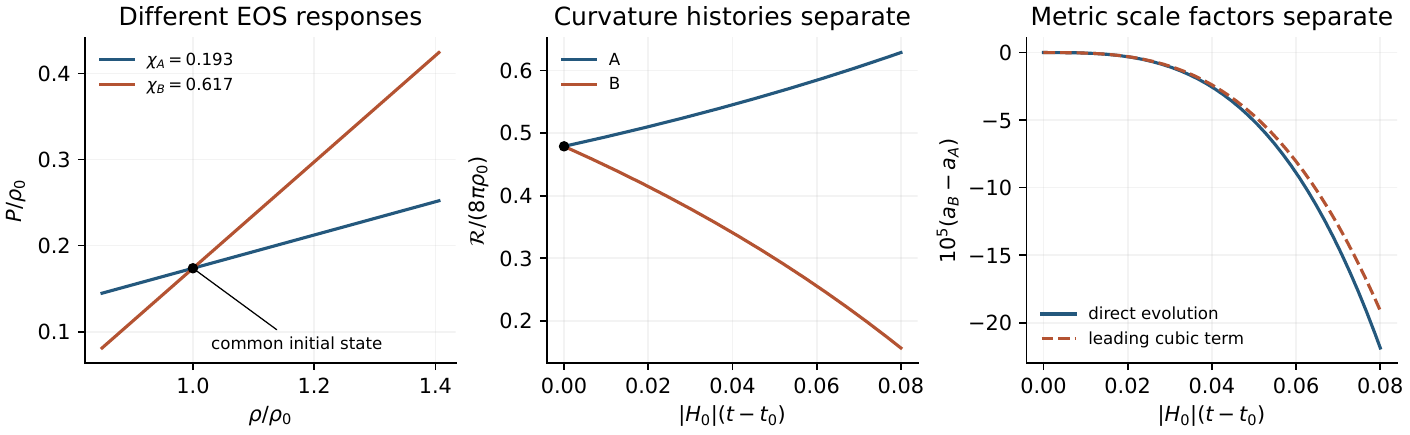}
\caption{MODEL: Same initial source and geometry, different subsequent evolution. Left: the two EOS curves cross at $(\rho_0,P_0)$. Center: the initially equal Ricci scalars separate. Right: the scale factors in Eq.~\eqref{eq:initialmetric} separate too; the dashed curve is the leading short-time cubic term of Eq.~\eqref{eq:metricdeparture}. At the right edge, this approximation differs from the direct scale-factor difference by approximately 13\%, reflecting higher-order time dependence rather than integration error. The full initial Riemann tensors agree by Eq.~\eqref{eq:initialcurvature}. This generic contracting example uses $\rho_0=0.731$, $P_0=0.127$, $\chi_A=0.193$, $\chi_B=0.617$, with $H_0=-\sqrt{8\pi\rho_0/3}$ in geometric units; the evolution remains inside the declared homogeneous-fluid domain.}
\label{fig:response}
\end{figure}

Figure~\ref{fig:response} places the response and geometric comparisons side by side. The common ratio $P_0/\rho_0$ fixes neither constitutive slope $dP_i/d\rho$. These are the two declared fluid models, not the two radiation gases of Sec.~3.

The comparison synchronizes proper time along comoving fluid worldlines on the initial slice. In particular, $\dot{\mathcal R}=u^a\nabla_a\mathcal R$ is a scalar derivative; its difference is not a coordinate effect. In the common spatial normalization of Eq.~\eqref{eq:initialmetric}, the metrics also separate. Writing $\tau=t-t_0$, their scale-factor difference begins at cubic order:
\begin{equation}
\Delta a(t_0+\tau)
=2\pi H_0(\rho_0+P_0)(\chi_B-\chi_A)\tau^3
+O(\tau^4).
\label{eq:metricdeparture}
\end{equation}
Equation~\eqref{eq:metricdeparture} expresses Visser's cosmological equation-of-state--jerk relation~\cite{visserjerk} as the difference between two scale-factor histories with matched initial data.
Their common initial metric, first derivative and curvature therefore do not fix the future metric without the material response. For unequal responses the curvature histories separate immediately, while the scale factors initially agree through quadratic order. Both models obey local stress--energy conservation.

Appendix~\ref{app:evolution} gives a first-law-consistent material energy and checks generic parameter choices for both signs of $H_0$; the figure shows one contracting example. Response observations can restrict models beyond the initial geometric data.

In a stellar setting, the relevant response need not follow the equilibrium composition. Reisenegger and Goldreich distinguish the derivative along a neutron star's chemical-equilibrium background from the response of a displaced element whose composition remains frozen. The resulting buoyancy supports core $g$-modes in their neutron-star model~\cite{reisenegger}. This is an established example of why the perturbative material constraints must be specified as well as the equilibrium profile.

Cosmological radiation gives an observational anchor. The same fermionic weighting used in Sec.~3 enters thermal radiation-density counting, while $N_{\rm eff}$ parametrizes the radiation abundance in cosmological inference. The physical thermal history also sets relative species temperatures, so it differs from the stipulated common-temperature mixture. A background EOS does not specify the perturbative stresses~\cite{hu}: free-streaming neutrinos generate acoustic phase shifts~\cite{bashinsky} and anisotropic stress that damps tensor modes~\cite{weinberg}. Follin et al. report a detection of the neutrino-induced phase shift in CMB data~\cite{follin}. This probes propagation beyond the background radiation density; it is a model-dependent example of response data restricting a material inference.

\begin{definition}[Material inventory accounting]\label{def:inventory}
An inventory account specifies currents $J_i^a$, reaction sources $\Gamma_i$ through $\nabla_aJ_i^a=\Gamma_i$, and fluxes through the boundary of the region being followed. For a spacelike slice with future-directed unit normal $n^a$ and induced proper volume element $dV$, its inventory is $N_i=-\int J_i^a n_a\,dV$. A linear combination is locally conserved when its combined reaction source vanishes; its inventory in an open region still changes by boundary flux. Conservation is imposed only where the material theory supplies it.
\end{definition}

The total-source equation and the constituent-current equations are different parts of the forward problem. Reactions can change species inventories while preserving specified combinations. Recovering such inventories requires those equations even when the geometry is known. An inventory here is a property of a specified modeled history, with conservation imposed only where its material theory supplies it; it is not an additional label assigned to the stationary black hole from its exterior parameters. This differs from assigning a black hole's baryon number by formation counting and then invoking conservation as its independent justification, the circular inference discussed by Bekenstein~\cite[p.~1245]{bekenstein72}.

\section{Exterior uniqueness and the constituent question}\label{sec:nohair}

Consider final exterior parameters without a record of the events that produced them. Could they distinguish matter present during formation from matter supplied by later accretion? The distinction is historical, whereas the final parameters describe a state. Measuring an accretion event can supply additional history; measuring only the final exterior is a different inverse problem.

The matched-parameter inventory example is already explicit in Wheeler's characterization reproduced by Teitelboim: black holes built from different baryon and lepton populations are compared at the same mass, charge and angular momentum~\cite[Fig.~1]{teitelboim}. We use this established distinction between exterior parameters and constituent history.

Bekenstein's static analysis gives a concrete example: within the exterior-field description considered there, external measurements do not assign an unambiguous baryon number~\cite{bekenstein72,bekensteinprl}. Hartle's neutrino-field analysis and Teitelboim's treatment of quantum-number measurability belong to the same early investigation of exterior attribution~\cite{hartle,teitelboim}. An exterior attribution result and a reaction or flux account of an inventory are distinct questions.

Electric charge has an exterior measurement through the flux of a long-range Maxwell field. Teitelboim contrasts this Gauss-law charge with massive-vector decoupling in his Schwarzschild calculation; Bekenstein excludes specified static massive fields under his coupling and regularity assumptions~\cite{teitelboim,bekenstein72}. Global baryon and lepton inventories have no analogous Maxwell flux in the declared theory; massive or confined interactions do not automatically provide such a long-range constituent-charge measurement. The reduction to $(M,J,Q)$ still requires the electrovacuum uniqueness hypotheses below: field range alone does not establish it. Exterior charge measurement and local inventory accounting therefore remain distinct.

Zurek and Thorne connect black hole entropy to the number of quantum-mechanically distinct ways a rotating, charged hole could have formed~\cite{zurekthorne}. Counting compatible formation ways and identifying which constituent history occurred are different inverse questions; here we ask whether the retained exterior data identify a specified history property.

For this application, let $\mathcal A_{\mathrm{hist}}$ contain solutions of specified GR--matter evolution models, with their initial data, environments and reaction and transport laws, that admit the chosen stationary exterior as an idealized late-time state or limit. The map $F_{\mathrm{ext}}$ retains only its $(M,J,Q)$, which fix the exterior metric within the chosen uniqueness class. The property $q$ may be an initial species inventory or the fraction of a tracked inventory delivered after a specified formation epoch. Its definition requires the model's current accounting. The ideal stationary limit is an assumption of this inverse problem, not a claim of exact settling at a finite time.

Here ``no-hair'' refers to stationary exterior uniqueness, not a general theorem of dynamical settling. We choose the electric-charge Kerr--Newman family with magnetic charge fixed to zero as our own restriction. We use a sufficient subclass of the uniqueness hypotheses in Ref.~\cite{uniqueness}: four-dimensional asymptotic flatness, an analytic stationary electrovacuum spacetime, a connected non-degenerate horizon, and $I^+$-regularity. Stationarity includes invariance of the electromagnetic field, $\mathcal L_k F_{ab}=0$, where $k^a$ is the stationary Killing field and $F_{ab}$ the Maxwell tensor. The non-degenerate restriction is stronger than the review's alternative of mean non-degeneracy or rotation. Within our class, $(M,J,Q)$ fixes the exterior. A dynamical approach to that family requires a separate evolution argument.

The stationary parameters do not specify initial data or time-dependent fluxes. If two histories in $\mathcal A_{\mathrm{hist}}$ have the same final parameters and different $q$, Criterion~\ref{crit:identifiability} makes $q$ non-identifiable from $F_{\mathrm{ext}}$. This states the inference test; the admissible range of histories must be established by the chosen evolution model.

The distinction also fixes the scope of the stellar comparison. An electrovacuum exterior has no material-fluid source to decompose locally. The question concerns the complete physical history compatible with that exterior, rather than particles occupying its vacuum region. No-hair constrains that inference without reconstructing the history.

For an observed candidate, a different class $\mathcal A_{\mathrm{cand}}$ compares physical models against actual measurements and their uncertainties. The distinction between evidence for a horizon and a proof from observations has long been emphasized in the mimicker literature~\cite{akl}. The response test parallels Sec.~\ref{sec:evolution}: retaining only initial background information can leave response laws undetermined, while measuring the response can distinguish models. In compact-object studies, models matching selected background parameters such as mass and spin can differ in tidal deformability, absorption and quasinormal-mode spectra~\cite{cardoso}. The background data being matched must be specified: agreement in $(M,J)$ is weaker than equality of a complete exterior metric. Adding response observables then refines the inverse set, just as it does for the fluid models.

For $F_{\mathrm{ext}}$, $\Pi$ retains only final stationary parameters. A time-resolved observation map can additionally retain radiation, accretion fluxes and memory. Gravitational memory relates lasting detector displacements to radiative flux~\cite{memory}; soft-hair proposals study further charges and soft degrees of freedom beyond the stationary labels~\cite{softhair}. These are distinct sources of history-sensitive information, not interchangeable with $(M,J,Q)$. Whether the enlarged observation set identifies a chosen constituent property still requires Criterion~\ref{crit:identifiability}. For an event-horizon model, the causal restriction on signals from the interior also remains part of the forward problem.

The scope of soft information also depends on the scattering description. Within their respective scattering assumptions, Mirbabayi and Porrati show that infrared dressing factorizes the soft dependence from the hard scattering; Bousso and Porrati obtain a corresponding decoupling of soft variables in their classical and quantum treatments~\cite{mirbabayi,boussoporrati}. These results do not identify soft charges with a complete constituent record.

Equality of stress--energy histories alone does not establish equality of the metric: gravitational initial and boundary data also matter. Once two models do share the same metric, up to the identification used to compare them, throughout the region determining an observable, any observation depending only on that metric with the same measurement prescription gives the same result. Further metric-dependent observations within that region cannot resolve the remaining material ambiguity. Material response can distinguish the models gravitationally if it changes the subsequent geometry within the observed domain. Non-gravitational channels, including electromagnetic emission, particle fluxes and direct material probes, require their own material predictions.

\section{First laws, fluxes and constituent history}

The local material law in Eq.~\eqref{eq:firstlaw} accounts for allowed changes of a specified physical state. The black hole law uses a different state space. For neighboring stationary electrovacuum black hole solutions, in units $G=c=\hbar=k_B=1$,
\begin{equation}
\delta M=\frac{\kappa}{8\pi}\delta A+\Omega_H\delta J+\Phi_H\delta Q
=T_H\delta S_{\mathrm{BH}}+\Omega_H\delta J+\Phi_H\delta Q.
\label{eq:bh}
\end{equation}
Here $M$ and $J$ are the asymptotic mass and angular momentum, $Q$ is electric charge, $A$ is horizon area, $\kappa$ is surface gravity, $\Omega_H$ is the horizon angular velocity relative to infinity, and $\Phi_H$ is the horizon electric potential relative to infinity. With the stationary time translation normalized at infinity, $T_H=\kappa/(2\pi)$ and $S_{\mathrm{BH}}=A/4$. Bekenstein introduced the area--entropy relation~\cite{bekenstein73}; Hawking's temperature supplies its standard normalization~\cite{hawking}. The mechanical and variational first laws are given in Refs.~\cite{bch,wald}.
This law relates variations along the specified stationary family; variational formulations also allow suitable nonstationary perturbations~\cite{wald}. The full stationary-system treatment of Bardeen, Carter and Hawking also retains external-matter contributions, and relativistic fluid first laws explicitly retain material variations~\cite{bch,iyer,friedman}. The displayed variables therefore depend on the system and the admissible variations being considered.

The mechanical law was derived from stationary spacetime structure and field equations, without requiring a constituent model of the matter that formed the hole~\cite{bch}. Hawking's radiation calculation gave the temperature identification its physical content~\cite{hawking}.

The connection to accreting matter is also established. The physical-process first law relates first-order changes in the black hole parameters to perturbing stress--energy and electric-current flux~\cite{waldbook,gaowald}. In the Gao--Wald treatment, the perturbing matter and current are initially localized away from the horizon and infinity and are assumed to fall entirely into an initially stationary electrovacuum hole, which settles to a final stationary electrovacuum state. It supplies a dynamical energy, angular-momentum and gauge-charge account within those assumptions. Relating a larger supplied inventory to the final parameter changes would also require accounting for any escaping flux.

In this physical-process account, energy is defined relative to the stationary time translation normalized at infinity, with rotation and electromagnetic work accounted for separately~\cite{gaowald,iyer}. It is not the local rest-frame material energy $E$ of Eq.~\eqref{eq:firstlaw}: the flux account includes the relation between local energy and asymptotic energy. Matter's contribution to the final mass does not make its baryon, lepton or species inventory an independently identifiable exterior parameter. Energy transfer and recovery of the constituent labels are different questions. Definitions~\ref{def:model} and~\ref{def:inventory} specify the additional information needed to follow those labels.

The material evolution connects local thermodynamic changes to asymptotic parameters through the coupled field equations and fluxes. Oppenheimer--Snyder collapse supplies a classic pressureless example of continued gravitational contraction~\cite{os}. A constituent-resolved account also follows species currents and reactions along the resulting history, including across trapped surfaces when present.

\section{Conclusion}

Two distinct inverse problems connect the examples in this note. In equal-source ambiguity, different material states produce the same total tensor on a common spacetime region. In projection ambiguity, different full descriptions share the retained observations. The first concerns realization of a source; the second concerns which information survives the observational map.

The gas example makes temperature and entropy differences explicit at equal source. The evolution example shows how a response measurement can distinguish models that share an initial source and geometry. For a stationary black hole exterior, the corresponding question concerns the histories compatible with its parameters, while the physical-process first law relates perturbing flux to parameter changes.

A material inference should specify the physical class, observation map and property being recovered, and test whether that property is constant across the compatible models. The complementary forward task is to determine which geometric descriptions the admitted material states and their evolution can realize. Material restrictions can establish a property or bound throughout that class without identifying a unique microscopic state. Observational compatibility, material realization and such class-wide bounds are distinct questions.

\section*{AI-tool use}
Agentic AI tools (Anthropic Claude and OpenAI Codex) assisted with numerical calculation checks, computational scripts, figure generation, and proofreading. The author takes responsibility for all content and conclusions.

\begingroup\small
\appendix
\section{Ideal-gas control and approximation ledger}\label{app:control}

For a massless species with degeneracy $g$, the equilibrium energy and number densities at zero chemical potential are
\begin{align}
\rho_{\pm}&=\frac{gT^4}{2\pi^2}\int_0^\infty\frac{x^3}{e^x\pm1}\,dx,\\
n_{\pm}&=\frac{gT^3}{2\pi^2}\int_0^\infty\frac{x^2}{e^x\pm1}\,dx.
\end{align}
The minus sign is bosonic and the plus sign fermionic. Relative to bosons, the fermionic energy and number integrals have factors $7/8$ and $3/4$. Thus
\begin{equation}
n_A=\frac{2\zeta(3)}{\pi^2}T_A^3,\qquad
n_B=\frac{[2+(3/4)6]\zeta(3)}{\pi^2}T_B^3,
\end{equation}
and Eq.~\eqref{eq:temperature} gives $n_B/n_A=(13/4)(8/29)^{3/4}\simeq1.237093$. These densities count the total ideal-gas populations; the zero-chemical-potential construction assigns no nonzero net neutrino lepton charge.

The script \emph{anc/verify\_example.py}, supplied in the accompanying ancillary directory, evaluates the four occupation integrals directly and checks the equal-source construction at three generic temperatures. It also integrates a common static spherical control volume in density and temperature variables. In geometric units the latter obeys
\begin{align}
m'&=4\pi r^2\rho,\\
\rho'&=-4\rho\frac{m+4\pi r^3\rho/3}{r(r-2m)},\\
(\ln T)'&=-\frac{m+4\pi r^3\rho/3}{r(r-2m)}.
\end{align}
The temperature equation is the Tolman--Ehrenfest relation~\cite{tolman}. Self-gravitating radiation in a box also has an established thermodynamic treatment: Sorkin, Wald and Zhang relate entropy extrema to static equilibria under their symmetry and local-equilibrium assumptions~\cite{swz}. MODEL: with $\rho_c=1$ and $R=0.18$, our integration gives $M(R)=0.018122730994$ and $2M/R=0.201363677714$. Tightening the relative tolerance from $10^{-10}$ to $10^{-12}$ changes $M$ by $1.6\times10^{-12}$; density and temperature formulations differ in density by at most $1.8\times10^{-12}$ relatively. This is an implementation check within their common EOS, since $\rho\propto T^4$. Assigning both gases the same temperature instead yields the unequal density ratio $29/8=3.625$, which the calculation detects.

The approximation ledger is: massless dispersion, ideal noninteracting gases, zero chemical potentials, a common four-velocity, specified isotropic thermal distributions, and externally specified boundary traction for the static control. For physical neutrinos the relativistic approximation requires thermal energies large compared with their rest energies. Finite masses and interaction corrections require revised material closure. The inverse-set criterion itself depends only on the declared forward map and admissible class.

\section{Conserved evolution with different material responses}\label{app:evolution}

At fixed entropy per particle, a conserved material density $n$ gives a first-law realization of Eq.~\eqref{eq:responsecontrol}:
\begin{equation}
\rho_i(n)=\frac{\rho_0+P_0}{1+\chi_i}
\left(\frac{n}{n_0}\right)^{1+\chi_i}
+\frac{\chi_i\rho_0-P_0}{1+\chi_i}.
\end{equation}
It satisfies $P_i=n\,d\rho_i/dn-\rho_i$, with $\rho_i(n_0)=\rho_0$ and $P_i(n_0)=P_0$. The fluid approximation is homogeneous, isotropic, adiabatic and barotropic; its linear response is specified locally, not inferred as a Standard-Model equation of state. Its flat FLRW gravitational data are a control problem rather than an isolated stellar configuration. The domain is restricted to positive density and enthalpy, the stated sound-speed range and the dominant-energy inequality.

For a numerical check, $\rho_i+P_i=(\rho_0+P_0)a^{-3(1+\chi_i)}$ gives an analytic density as a function of scale factor. The accompanying \emph{anc/verify\_evolution.py} compares direct evolution with this relation, checks Eq.~\eqref{eq:curvaturedeparture} at generic states for expansion and contraction, and tests the equal-response null. The additional \emph{anc/verify\_metric\_control.py} reconstructs the FLRW curvature from the metric, verifies the initial-curvature and cubic-departure expressions symbolically, and checks the short-time scalings by direct evolution at three generic states for both signs of $H_0$. Time and integration refinement test numerical error. These checks share the declared fluid model and certify its implementation, not microscopic accuracy or a compact-object endpoint.

\endgroup
\begingroup
\small
\setlength{\parskip}{2pt}

\endgroup
\end{document}